\documentclass[aps,prd,preprint,amsmath,amssymb,showpacs]{revtex4}

\usepackage{bm}
\usepackage{enumerate}

\def\J{\mathbf{J}}
\def\Q{\mathbf{Q}}
\def\T{\mathbf{\Theta}}
\def\B{\mathbf{B}}
\def\E{\mathbf{E}}
\def\eps{\boldsymbol{\epsilon}}

\def\beq{\begin{equation}}
\def\eeq{\end{equation}}
\begin{document}
\title{Noether charges and black hole mechanics in Einstein--{\ae}ther 
theory}
\author{Brendan Z. Foster}
\email[]{bzf@umd.edu}

\affiliation{Dept.~of Physics, University of Maryland, College
Park, MD 20742-4111, USA}

\date{December 9, 2005}
%
%
\begin{abstract}
The Noether charge method for defining the Hamiltonian
of a diffeomorphism-invariant field theory is applied to 
``Einstein--{\ae}ther" theory, in which gravity
couples to a dynamical, timelike, unit-norm vector field.  Using
the method, expressions are obtained for the total energy, momentum, and 
angular momentum of an Einstein--{\ae}ther space-time.  
The method is also used to discuss
the mechanics of Einstein--{\ae}ther black holes.  
The derivation of Wald, and Iyer and Wald, of the first law of 
black hole thermodynamics fails for this theory, because the unit vector
is necessarily singular at the bifurcation surface of the Killing horizon.
A general identity relating variations of energy and angular
momentum to a surface integral at the horizon
is obtained, but a thermodynamic interpretation,
including a definitive expression for the black hole
entropy, is not found. 
\end{abstract}
\maketitle
%
%
%
\section{Introduction}

`Einstein--{\ae}ther' theory, or `{\AE}-theory' for 
short, is a 
vector tensor theory in which the vector field, or `{\ae}ther', is 
constrained to be everywhere timelike and of fixed norm.  This theory has 
received increasing attention lately, stemming from a broader interest in 
the possibility that Lorentz symmetry is not an exact symmetry of nature. 
The vector field defines a ``preferred" frame, thus allowing for 
violations of local Lorentz symmetry, while its status as a dynamical 
field preserves diffeomorphism invariance. The condition on the vector 
norm (which can always be scaled to unity) ensures that the {\ae}ther just 
picks out a timelike direction, and is required for stability of the 
theory at the semi-classical level~\cite{Elliott:2005va}.
For a review of properties of this theory and references to earlier
work, see~\cite{Eling:2004dk}. 

The conventional {\AE}-theory action, defined below, contains four free
parameters.  Constraints on the acceptable values of the parameters are
implied by observational evidence, but 
one can also argue for limits
imposed by theoretical considerations (for a summary of the known 
constraints, see~\cite{Foster:2005dk}).  A possible requirement that
motivates the present work is that the theory should satisfy some form of
energy positivity. It may be that imposing positivity for all 
solutions is
more restrictive than necessary, or perhaps that one should only require
positivity in the rest-frame of the {\ae}ther.  Whatever the argument, an
expression for the energy is required to know how the parameters are
constrained.

With this goal in mind, I give here an expression for the
total energy of an asymptotically-flat {\AE}-theory space-time,
as well as expressions for the total momentum and angular momentum.
These 
are generated via the `Noether charge'
method~\cite{Wald:1993nt,Iyer:1994ys}
of defining the value of the on-shell Hamiltonian
for a diffeomorphism-invariant field theory, directly from the
theory's Lagrangian.  The conventional ADM and Komar expressions,
which have the form of integrals at spatial infinity, acquire 
{\ae}ther-dependent corrections due to the non-vanishing of the 
{\ae}ther at infinity.  Parameter constraints are not discussed. 
The results 
here complement those of
Eling~\cite{Eling:2005zq}, in which expressions for the total
energy and the energy of linearized wave-modes are derived via
pseudotensor methods.  

The Noether-charge method also allows one to write down a differential 
identity that governs variations of stationary, axi-symmetric black hole 
solutions.  As shown by Wald~\cite{Wald:1993nt} and Iyer and 
Wald~\cite{Iyer:1994ys}, in a wide variety of theories this identity can 
be massaged into the familiar form of the `first law' of black hole 
mechanics and then interpreted as a law of thermodynamics.  The recent 
discovery of {\AE}-theory black hole solutions~\cite{TEDCHRIS} motivates 
the study of the first law for {\AE}-theory black holes.  
The authors of~\cite{TEDCHRIS} demonstrate existence of these solutions, 
but 
have not found
analytic expressions for the fields; therefore, the form of the first 
law cannot be inferred directly from the solutions.  One can, however, 
attempt to derive the law via the Noether charge method.  Unfortunately,
the algorithms
of~\cite{Wald:1993nt,Iyer:1994ys} fail for {\AE}-theory 
since the vector field cannot be regular on the bifurcation surface of the 
horizon, where a crucial calculation is performed.  Below, 
a law resembling the first law is derived by less elegant means for 
static, 
spherically-symmetric solutions, but a thermodynamic interpretation of 
this expression is not given. In particular, a definitive expression for 
the horizon entropy in {\AE}-theory has not yet been found.

The Noether charge methodology is briefly reviewed in Section~\ref{2nd}.
The requisite differential forms for {\AE}-theory are derived in
Section~\ref{3rd}.  These are used to determine expressions for the total
energy, momentum, and angular momentum of an asymptotically-flat
{\AE}-theory space-time in Section~\ref{4th}.  The first law of 
{\AE}-theory
black holes is discussed in Section~\ref{5th}.  I follow the conventions
of~\cite{WALDBOOK}, except that the metric here has signature
$({+} {-}{-} {-})$. I employ units in which $\hbar=c=1$.

%
%
\section{Noether Charge Methodology}
\label{2nd}

I will summarize here the application of the Noether charge
method~\cite{Wald:1993nt,Iyer:1994ys} to the definition of total
energy, momentum, and angular momentum of an asymptotically-flat
space-time.  (We take the space-time to be four-dimensional, but
the method can be applied in any dimension.) Given a 
diffeomorphism-invariant
field theory defined from an action principle, one can
construct a phase space with symplectic structure from the space
of field configurations and the theory's Lagrangian. For the case
of an {\AE}-theory system on a globally hyperbolic space-time, the
phase space structure permits a well-defined Hamiltonian
formulation. For every diffeomorphism on space-time, generated by
vector field $\xi^a$, there is a corresponding evolution in phase
space, with Hamiltonian generator $H_{\xi}$. This generator is
implicitly defined through Hamilton's equation, which takes the
form~\cite{Wald:1993nt,Iyer:1994ys}
\begin{equation}\label{HAMEQ}
    \delta H = \int_C (\delta\J - d(i_{\xi}\T))
\end{equation}
where $\J$ and $\T$ are differential 3-forms that depend on the
dynamical fields, a variation of the fields, and the vector field
$\xi^a$; the surface of integration is a space-like Cauchy surface
$C$ of the space-time.

The forms $\T$ and $\J$ are obtained from the theory's Lagrangian
as follows. Let the Lagrangian $\mathbf{L}$ be a four-form
constructed locally out of the dynamical fields, denoted
collectively by $\psi$. The 3-form $\T$ is defined by the
variation of $\mathbf{L}$ due to a variation of $\psi$:
\begin{equation}\label{VARI}
    \delta \mathbf{L} = \mathbf{E}[\psi]\cdot\delta\psi
            + d\T[\delta\psi],
\end{equation}
where $\E[\psi]$ are identified as the equations of motion for the
fields, the dot representing contraction over appropriate indices.

This definition only determines $\T$ up to the addition of a
closed form, which must be exact by the result of~\cite{Wald1990}.
The contribution to
$\delta H$~\eqref{HAMEQ} from an asymptotic boundary is typically
not effected by such ambiguity, though, since the fall-off conditions on
the dynamical fields that guarantee convergence of $\delta H$
imply that any covariant, exact three-form added to $\T$ will give
no asymptotic contribution to $\delta H$.  Such is the case for
{\AE}-theory with the conditions chosen below.  A contribution
might arise given an inner boundary to the space-time.  Here
(Sec.~\ref{5th}) we only consider stationary configurations on
such space-times, and one can then show that the contribution to
$\delta H$ vanishes.  We will therefore fix the definition of $\T$
by taking the ``most-obvious" choice that emerges from variation
of the Lagrangian.

To each vector field on space-time $\xi^a$, we associate the
Noether current 3-form $\J[\xi]$,
\begin{equation}\label{JJJ}
    \J[\xi] = \T[\mathcal{L}_\xi\psi] - i_{\xi} \mathbf{L}.
\end{equation}
This current is conserved, $d\J = 0$, for arbitrary $\xi^a$ when
$\psi$ satisfies the equations of motion. This fact
implies~\cite{Wald1990} that $\J$ can be expressed in the form
\begin{equation}\label{QQQ}
    \J[\xi] = d\Q[\xi]
\end{equation}
when $\E[\psi] = 0$.
If in addition $\delta \psi$ is such that the equations of motion
linearized about $\psi$ are satisfied, then $\delta \J = d\delta
\Q$, where here and below we choose $\delta \xi^a = 0$.  $\Q$ is
only defined up to addition of a closed, hence
exact~\cite{Wald1990}, 2-form, but this ambiguity does not
effect $\delta H$.  We will therefore fix the definition of $\Q$
by taking the ``most-obvious" choice.

An additional ambiguity can arise if one thinks of the Lagrangian
$\mathbf{L}$ as defined only up to the addition of a boundary term,
i.e.~an exact 4-form.
Adding such a form to $\mathbf{L}$
effects $\T$, $\J$, and $\Q$ individually but leads to no net
effect on $\delta H$. We will fix this ambiguity by again taking
the ``most-obvious" choices for the forms.

The Hamiltonian differential evaluated on-shell---when the full
and linearized equations of motion are satisfied---is thus a
surface term
\begin{equation}
    \delta H_{\xi}= \int_{\partial C}(\delta\Q - i_{\xi}\T).
\end{equation}
We will restrict attention to the case where $C$ is asymptotically
flat at spatial infinity.  The boundary of $C$ will consist of a
surface ``at infinity''---the limit of a two-sphere whose radius
is taken to infinity---and a possible inner surface, such as a black hole
horizon.

One can define a Hamiltonian function $H_{\xi}$ if there exists a
2-form $\B$ such that
\begin{equation}
      \int_{\partial C}  \delta(i_\xi \B) = \int_{\partial C}
    i_{\xi} \T.
\end{equation}
We then define the Hamiltonian as
\begin{equation}
    H_{\xi} = \int_{\partial C} (\Q - i_\xi\B).
\eeq
We can assume that the fall-off conditions on the fields are such
that at infinity, $d(\Q - i_\xi\B) = 0$.  It follows that the
value of the contribution to $H_{\xi}$ from the surface at
infinity is conserved and can be interpreted as the conserved
quantity associated with the symmetry generated by $\xi^a$.

One can define the total energy $\mathcal{E}$ of the
space-time to be the value of the asymptotic Hamiltonian for the
case where $\xi^a$ is a time translation $t^a = (\partial/\partial t)^a$ 
at infinity
\begin{equation}\label{NRG}
    \mathcal{E} = \int_{\infty}(\Q[t]- i_t\B).
\end{equation}
Likewise, with
$x_i^a=(\partial/\partial 
x_i)^a$ a constant, spatial translation at infinity, one can define
the total momentum in the $x^a_i$-direction $\mathcal{P}_i$ as
\beq
    \mathcal{P}_i = -\int_{\infty} (\Q[x_i] - i_{x_i}\B).
\eeq
One can define the total angular momentum
$\mathcal{J}$ (about a particular axis) via a vector field 
$\varphi^a$ 
that 
is a rotation at
infinity, tangent to the bounding 2-sphere.  The pull-back to the
boundary of $i_\varphi \B$ vanishes, giving
\begin{equation}\label{ANG}
    \mathcal{J} = - \int_{\infty}\Q[\varphi].
\end{equation}

I note parenthetically that it follows from this definition that
the total angular momentum must be zero for any axi-symmetric
configuration (one for which $\mathcal{L}_{\varphi} \psi = 0$), on
$C$ possessing no inner boundary.  This follows from the vanishing
of $\J[\varphi] = d\Q[\varphi]$, when evaluated on such a
configuration and pulled back to $C$.  This result does not appear
to have been stated explicitly with this generality before,
although an early application is found in the proof of Cohen and
Wald~\cite{COHEN} that there are no rotating, axi-symmetric geons,
in work that predates the precise formulation of the Noether
charge method. This result also provides a short proof that there
can be no rotating, axi-symmetric boson stars in general
relativity.
This generalizes the known
result~\cite{Yoshida:1997qf} that are no \emph{stationary}, rotating,
axi-symmetric boson stars.  (Here, axi-symmetry must include any complex
argument of the scalar field, as well as its modulus; this is a 
stronger sense of `axi-symmetric' than is common in the 
boson-star-related
literature.)   In the presence of an inner boundary, such as
an event horizon, the vanishing of $d\Q$ implies that the total
angular momentum, i.e.~the integral of $\Q$ over the boundary at
infinity, is equal to the integral of $\Q$ over the inner
boundary. Consequently, this result is not in conflict with the
existence of rotating, axi-symmetric black holes.

%
%
\section{{\AE}-theory Forms}
\label{3rd}

In this section, we will give the explicit expressions  of the
differential forms defined above, for {\AE}-theory. The conventional,
second-order {\AE}-theory Lagrangian 4-form $\mathbf{L}$ is
\begin{equation} \label{LAG}
\begin{split}
    \mathbf{L} = \frac{-1}{16\pi G}\bigl(R&
    + c_1 (\nabla_a u_b) (\nabla^a u^b)
    + c_2 (\nabla_a u^a)( \nabla_b u^b)\\ &
    + c_3 (\nabla_a u^b) (\nabla_b u^a)
    + c_4 (u^a\nabla_a u^c) (u^b \nabla_b u_c)\bigr) \eps
\end{split}
\end{equation}
where $R$ is the scalar curvature of the metric $g_{ab}$, the
$c_i$ are constants, and $\eps$ is the canonical volume form
associated with $g_{ab}$. This Lagrangian is the most general
(modulo a boundary term) that is covariant, second-order in
derivatives, and consistent with the constraint $u^a u_a =
1$.
The constraint can be accounted for by adding to
$\mathbf{L}$ a term of the form
\beq
    \lambda\bigl(u^a u^b g_{ab} - 1\bigr)\eps
\eeq
where $\lambda$ is a Lagrange multiplier.

Varying $\mathbf{L}$, we obtain $\T$:
\begin{multline}\label{THETA}
    \T_{abc} = \frac{1}{16\pi G}\eps_{dabc}
        \Biggl[g^{de}g^{fh}
        \left(\nabla_e\delta g_{fh}
                -\nabla_f\delta g_{eh}\right)\\
        -\biggl(2\, K^{d}_{\phantom{d}e}\delta u^e
        +\Bigl(K^{ef}u^d
        + \bigl(K^{df} - K^{fd}\bigr)u^e
        \Bigr)\delta g_{ef}\biggr)\Biggr],
\end{multline}
where
\begin{equation}
    K^a_{\phantom{i}m} = (c_1 g^{ab}g_{mn} + c_2 \delta^a_m \delta^b_n +
            c_3 \delta^a_n \delta^b_m + c^4 u^a u^b g_{mn})\nabla_b u^n.
\end{equation}
From this follows $\J$~\eqref{JJJ}:
\begin{equation}\label{JNOETH}
    \J_{abc} = \frac{1}{16\pi G}\eps_{dabc}
            \left(A^{def}_{\phantom{def}h}
                \nabla_{e}\nabla_{f}\xi^h
            + B^{de}_{\phantom{de}h}\nabla_e\xi^h
            + C^d_{\phantom{d}h}\xi^h\right),
\end{equation}
where
\begin{subequations}
\begin{align}
    A^{def}_{\phantom{def}h} &=
        \bigl(-g^{ef}\delta^d_h + g^{d(e}\delta^{f)}_{\smash[t]{h}}\bigr),\\
    B^{de}_{\phantom{de}h} &= 2\bigl(
            K^{\smash{[d}}_{\phantom{[d}h}u^{e]}+K_h^{\phantom{h}\smash{[d}}u^{e]}
            +K^{[ed]}u_h\bigr),
\end{align}
\end{subequations}
and we will not need the form of $C^d_{\phantom{d}h}$.

The Noether charge $\Q$~\eqref{QQQ} can be extracted via an
algorithm of Wald~\cite{Wald1990}, yielding
\begin{equation}\label{QNOETH}
\begin{split}
    \Q_{ab} &= \frac{1}{16\pi G}\,\eps_{abcd}\Bigl[
    \frac{2}{3}A^{cdf}_{\phantom{def}h}\nabla_f\xi^h
        + \frac{1}{2}B^{cd}_{\phantom{dc}h}\xi^h\Bigr]\\
        &=\frac{1}{16\pi G}\,\eps_{abcd}\Bigl[\nabla^c\xi^d
        +
        \Bigl(\bigl(K^c_{\phantom{c}h}+K_h^{\phantom{h}c}\bigr)u^d
            -K^{cd}u_h\Bigr)\xi^h\Bigr].
\end{split}
\end{equation}
%

%
%
\section{Conserved Quantities}
\label{4th}

We now consider the expressions for the total energy, momentum,
and angular momentum of an asymptotically flat space-time in
{\AE}-theory. For the requisite integrals to be convergent, we must
define fall-off conditions on the fields and their variations.  We
will assume that at spatial infinity, there exists an asymptotic
Cartesian coordinate basis, with respect to which the components
of the metric and its derivatives are
\beq
    g_{\mu\nu}=\eta_{\mu\nu} + O(1/r),
\eeq
and
\beq
        \frac{\partial g_{\mu\nu}}{\partial x^\alpha} =
        O(1/r^2),
\eeq
where $\eta_{ab}$ is the flat metric.  The variations of the
metric $\delta g_{ab}$ must be $O(1/r)$.  For the {\ae}ther,
we require that
\beq
    u^\mu = \bar{u}^\mu + O(1/r),
\eeq
where asymptotically, $\nabla_a {\bar u}^b = 0$.
We can always choose to effect an asymptotic
Lorentz boost so that ${\bar u}^a = t^a$ at infinity, i.e.~we are
in the rest-frame of the {\ae}ther. With respect to the asymptotic
Cartesian basis,
\beq
    \frac{\partial u^\mu}{\partial x^\alpha} = O(1/r^2).
\eeq
The variation $\delta u^a$ will be assumed to be $O(1/r)$.

We turn now to the total energy.  One finds in this case that with
the above fall-off conditions, $\T = \T_{G} + O(1/r^3)$
asymptotically, where $\T_{G}$ is the form which arises for GR in
vacuum. Hence, we can choose $\B = \B_{G}$, the vacuum form.  The total 
energy can then be written as
$\mathcal{E} = \mathcal{E}_{G} + \mathcal{E}_{\AE}$,
where $\mathcal{E}_{G}$ is the standard ADM mass~\cite{WALDBOOK}
\beq
    \mathcal{E}_{G} = \frac{1}{16\pi G}\sum_{i,j=1}^3\int_\infty dS\;
                        r^i (\partial_i g_{jj}
                        - \partial_j g_{ij}),
\end{equation}
where $dS$ is the spherical area element and $r^a =
(\partial/\partial r)^a$.  The
{\ae}ther portion $\mathcal{E}_{\AE}$ is
\begin{equation}
\begin{split}
    \mathcal{E}_{\AE}
                &=\frac{1}{16\pi G}\int_\infty
                        dS\;2\,r_{[c}t_{d]}
                            \Bigl(
                            \bigl(K^{ct}+K^{tc}\bigr)\bar{u}^d
                            -K^{cd}\bar{u}^t\Bigr)
                            \\
                        &= \frac{1}{8\pi G}\int_\infty dS\;
		t_a(r^b t_c + r_c t^b)K^a_{\phantom{a}b}{\bar u}^c.
\end{split}
\end{equation}
Setting ${\bar u}^a = t^a$ gives
\beq
\begin{split}
    \mathcal{E}_{\AE} &= \frac{(c_1 + c_4)}{8 \pi G}\int_\infty dS\;t^a 
		r_b\nabla_a u^b\\
                &= \frac{(c_1 + c_4)}{8\pi G}\int_\infty dS\;
    \bigl( \partial_t u^r + \partial_r u^t\bigr),
\end{split}
\eeq
where in the first line we have used the fact that $t_a \nabla_b u^a = u_a 
\nabla_b
u^a + O(1/r^3)$, and in the second we have used the constraint, which 
requires
$\partial_\mu u^t = -(1/2)\partial_\mu g_{tt} 
+O(1/r^3)$.

This expression can be evaluated more explicitly for a static,
spherically-symmetric, matter-free solution, as the
asymptotic forms of the fields are known~\cite{Eling:2003rd}. In
isotropic coordinates, the line element has the form:
\begin{equation}
    dS^2 = N(r)dt^2 - B(r)(dr^2 + r^2d\Omega)
\end{equation}
and $\bar u^a = t^a$. Assuming the generic case $c_1+c_2+c_3 \neq
0$, one finds that to $O(1/r)$, $N = 1- (r_0/r)
\text{ and } B = 1 + (r_0/r)$, for arbitrary constant $r_0$. The
total energy is then
\begin{equation}
    \mathcal{E} = \frac{r_0}{2G}(1-\frac{c_1+c_4}{2}).
\end{equation}
This result was previously found by Eling using pseudotensor
methods~\cite{Eling:2005zq}. The quantity
\beq\label{GNEWT}
    G_N = G(1-\frac{(c_1 + c_4)}{2})^{-1}
\eeq
has been identified in studies of the {\AE}-theory Newtonian
limit~\cite{Foster:2005dk,Carroll:2004ai} as the 
value of Newton's constant that
one would measure far from gravitating matter (assuming no direct
interaction between {\ae}ther and non-{\ae}ther matter). We can
define a Newtonian gravitating mass $M = 2r_0/G_N$, in which case
\beq\label{EEQM}
    \mathcal{E} = M.
\eeq

The total momentum in the $x_i^a$ direction also has the form
$(\mathcal{P}_{G})_i + (\mathcal{P}_{\AE})_i$, where $(\mathcal{P}_{G})_i$ 
is
the standard ADM momentum~\cite{WALDBOOK},
\beq
    (\mathcal{P}_{G})_i = \frac{1}{16\pi G}\sum_{j,k=1}^3 \int dS\;
	r^j\big(\partial_0 g_{ji} - \partial_j g_{0i} +
		\delta_{ij}(\partial_k g_{0k}
			-\partial_0 g_{kk})\big).
\eeq
The {\ae}ther contribution, setting $\bar u^a = t^a$, is
\begin{equation}
    (\mathcal{P}_{\AE})_i = \frac{-1}{16 \pi G} \int_{\infty} dS \;
          (c_1 + c_3)(r^a\nabla_a u_i + r_a \nabla_i u^a)
                                +2c_2 r_i \nabla_b u^b.
\end{equation}

The total angular momentum takes the form $\mathcal{J}_{G} +
\mathcal{J}_{\AE}$, where $\mathcal{J}_{G}$ is the generalization
to a non-axi-symmetric space-time of the conventional Komar
expression for vacuum GR~\cite{WALDBOOK},
\beq
    \mathcal{J}_G = \frac{1}{16\pi G}\int_{\infty} dS\;
    n_{ab}\nabla^a\varphi^b,
\eeq
where $n_{ab}$ is the binormal of the boundary of $C$.
The {\ae}ther contribution $\mathcal{J}_{\AE}$ is
\begin{equation}
    \mathcal{J}_{\AE} = \frac{-(c_1 + c_3)}{16 \pi G}
                                    \int_{\infty} dS \;
                                        2\,r_{(a}\phi_{b)}\nabla^a
                                        u^b,
\end{equation}
having set $\bar u^a = t^a$.

%
%
\section{First Law of Black Hole Mechanics}
\label{5th}

For a stationary black hole space-time, the Noether charge
formalism allows one to write down a differential identity that
relates variations in the total energy and angular momentum to
variations of integrals over a cross-section of the horizon.  It
has been shown~\cite{Wald:1993nt,Iyer:1994ys} that this identity
becomes the `first law' of black hole mechanics/thermodynamics
for a wide variety of generally covariant gravitational theories.

A one-parameter family, for a fixed set of $c_i$ values, of static, 
spherically-symmetric {\AE}-theory black hole solutions has been
shown to exist\cite{TEDCHRIS}.  The existence proof found 
in~\cite{TEDCHRIS} is based on numerical integration of the field 
equations, and analytic expressions for the fields are only known 
asymptotically (see~\cite{Eling:2003rd}).  
Thus, we cannot obtain a first law by directly examining the solutions.
We can instead apply the Noether charge method and attempt to 
massage the variational identity into a form resembling
the familiar first law.  The closest we will come here will
be for the static, spherically-symmetric case, for which we will 
show that the identity can be written in the form
\beq\label{LAW}
    \delta M = \frac{\kappa}{8\pi G}\big[(1 + \phi N) \delta A
            + \phi A\delta O\big],
\eeq
where $M$ is the black hole mass (i.e.~the total energy,
c.f.~\eqref{EEQM}), $A$ is the area of the horizon, $N$ and $O$ are
quantities depending on the metric, {\ae}ther, and the local
geometry of the horizon, and $\kappa$ and $\phi$ are parameters
defined below. Although this expression resembles the familiar
first law, it does not lead to an obvious thermodynamic
interpretation; in particular, we do not obtain a definitive
expression for the horizon entropy.  

The variational identity of interest is derived via the Noether
charge method by applying Hamilton's equation~\eqref{HAMEQ} to
perturbations of an asymptotically flat, stationary, axi-symmetric
configuration containing a Killing horizon.  A Killing horizon is
a null hypersurface to which a Killing field is normal---we take
it to define the black hole horizon. The Cauchy surface $C$ is
assigned a boundary consisting of the 2-sphere ``at infinity'' and
the surface $B$ where $C$ meets the horizon $\cal H$.  We will
assume that $B$ is compact.  Choose $\xi^a$ to be the
horizon-normal Killing field $\chi^a$, normalized as
\beq
    \chi^a = t^a + \Omega \phi^a,
\eeq
where $t^a$ is the stationary Killing field with unit norm at
infinity, and $\phi^a$ is the axi-symmetric Killing field; the
constant $\Omega$ defines the angular velocity of the horizon. As
$\delta\J[\chi] - d(i_\chi \T)$ is linear in $\mathcal{L}_\chi
\psi = 0$, $\delta H_\chi$ vanishes. From the definitions
of the total energy~\eqref{NRG} and angular momentum~\eqref{ANG},
the identity emerges:
\beq\label{VARID}
    \delta\mathcal{E} - \Omega\delta\mathcal{J} = \int_{B} \delta
    \Q - i_\chi \T.
\eeq
Note that the vanishing of $\delta\J[\chi] - d(i_\chi \T)$ also
implies that the choice of $B$ is arbitrary.

There is no precise definition of a ``first-law form" of an
expression; roughly speaking, however, by analogy with the
conventional thermodynamic expression, a black hole first law
should relate variations of ``macroscopic" variables---global
variables and other parameters that describe the black hole
space-time. Considering the explicit form that the identity takes
for {\AE}-theory, where $\T$~\eqref{THETA} and $\Q$~\eqref{QNOETH}
are as defined above, we can see that further manipulation is
required for the identity~\eqref{VARID} to take a first-law form.

An algorithm for massaging~\eqref{VARID} into such a form and
defining the entropy associated with the horizon was given by
Wald~\cite{Wald:1993nt} and improved upon by 
Iyer and Wald~\cite{Iyer:1994ys}. For the algorithm to apply, it is
necessary that the stationary space-time be extendible to one
whose Killing horizon possesses a bifurcation surface---a
cross-section on which the horizon-normal Killing field
vanishes---on which all dynamical fields are regular.  In that
case, one can work with the extended space-time and choose $B$ to
be the bifurcation surface. The algorithm relies on the universal
behavior of $\chi^a$ in a neighborhood of the bifurcation surface
and reduces the horizon terms to the form $(\kappa/2 \pi G) \delta
S$.  Here, $\kappa$ is the `surface gravity' of the horizon,
defined by $\kappa^2 = -\frac{1}{2}\nabla_a \chi_b \nabla^a
\chi^b$, evaluated on the bifurcation surface, and
\beq
    S = 2\pi \int_{B} E^{abcd}n_{ab}n_{cd},
\eeq
where $n_{ab}$ is the binormal of $B$, and $E^{abcd}$ is the
functional derivative of the Lagrangian with respect to the
Riemann tensor $R_{abcd}$, treating it as a field independent of
the metric. General kinematical arguments~\cite{Kay:1988mu} in the
context of quantum field theory in curved space-time indicate that
the temperature due to thermal radiation associated with a Killing
horizon is always $\kappa/2 \pi$.  The form of the horizon terms
then suggest that we identify $S/G$ as the thermodynamic entropy
associated with the horizon.

Unfortunately, the above requirement cannot be met for any
{\AE}-theory configuration~\cite{Eling:2004dk}.  Racz and
Wald~\cite{Racz:1995nh} have shown that a space-time containing a
Killing horizon can be extended smoothly to one containing a
bifurcation surface if the horizon has compact cross-sections and
constant, non-vanishing surface gravity.  Regular extensions of
matter fields on that space-time are not guaranteed. In fact, no
such extension can exist for a vector field $u^a$ that is
invariant under the Killing flow and not tangent to a horizon
cross-section. The Killing flow acts at the bifurcation surface as
a ``radially-directed" Lorentz boost, under which only vectors
tangent to the surface can be invariant. In particular, the
{\ae}ther cannot possess a regular extension, since it is
constrained to be time-like, while a cross-section of a null
surface must be space-like.

We can only proceed by less elegant and less general means.  We will
now restrict attention to the case of a perturbation between
spherically-symmetric, static solutions, and show that in this case the 
variational
identity~\eqref{VARID} can be written in the form~\eqref{LAW}.

Consider a variation between static, spherically-symmetric
solutions, each containing a Killing horizon $\mathcal{H}$.
Identify the solutions such that the horizons coincide, and so
that the Killing orbits coincide in a neighborhood of the horizon.
That this can be done follows from the construction of
``Kruskal-like" coordinates in ref.~\cite{Racz:1995nh}.  We can
further define these coordinates so that the variations of the
non-angular components of the metric vanish on $\cal H$: the line
element near $\cal H$ takes the form
\beq
    ds^2 = G dU dV - R^2 d\Omega^2.
\eeq
where $G$ and $R$ are functions of the quantity $UV$, and $\cal H$
is defined by $UV = 0$; we then effect a rescaling of $U$ and $V$
such that $G(0) = 1$ for each solution.

Another effect of this identification~\cite{Jacobson:1993vj} is
that near $\cal H$, the Killing vector $\chi^a$ with surface
gravity $\kappa_0$ in the unperturbed solution coincides with the
Killing vector with the same surface gravity $\kappa_0$ in the
perturbed solution. From this fact, it follows that on $\cal H$,
$\delta \Q[\chi] = \kappa \delta\Q[k]$, where $k^a =
\kappa^{-1}\chi^a$ is the unit-surface-gravity Killing field near
$\cal H$ for both configurations, and is held fixed in the
variation of $\delta\Q[k]$.

We will consider the portion of $\cal H$ defined by $U=0, V
> 0$, and a cross-section $B$ corresponding to some value of $V$. We can
define a null dyad on $\cal H$ consisting of $k^a$ and $\bar k^a$,
where $\bar k^a$ is the unique null vector normal to $B$ such that
$k_a \bar k^a = 1$. From the vanishing on $\cal H$ of the
variations of $k^a$ and the transverse components of the metric,
it follows that $\bar k^a$ is the same vector field for both
solutions; i.e.~$\delta \bar k^a = 0$. The metric $h_{ab} =
-g_{ab}+2k_{(a}\bar k_{b)}$ induced on $B$ has a variation
\beq
    \delta h_{ab} = \frac{\delta A}{A} h_{ab}
\eeq
where $A$ is the area of $B$. The {\ae}ther $u^a$ can be decomposed
with respect to this dyad:
\beq
    u^a = \frac{1}{2\phi} k^a + \phi \bar k^a.
\eeq
where $\phi = u^a k_a$.  We then have
\beq
    \delta u^a =
    -\frac{\delta\phi}{\phi}(\frac{1}{2\phi}k^a-\phi \bar k^a) \equiv
    -\frac{\delta\phi}{\phi}\bar u^a.
\eeq
A non-null dyad normal to $B$ consists of $u^a$ and the orthogonal
unit-vector $\bar u^a$. The bi-normal $n_{ab}$ of $B$ can be
expressed in various ways:
\beq
    n_{ab} = 2\bar k_{[a}k_{b]} = 2 u_{[a}\bar u_{b]}
        = \frac{2}{\phi} k_{[a}\bar u_{b]}.
\eeq

Now, the algorithm cited above can be used to evaluate the \ae
ther-independent horizon terms, which
give~\cite{Wald:1993nt,Iyer:1994ys} the standard contribution
$(\kappa/8\pi G)\delta A$. Evaluating the \ae ther-dependent
portion of $\Q[k]$, pulled-back to $B$, we find that
\beq
\begin{split}
    \Q_{\AE}[k] &= \frac{1}{16\pi G}\eps
            n_{ab}k^c\bigl(u^a(K^b_{\phantom{b}c} + K_c^{\phantom{c}b})
            + u_c K^{ab}\bigr)\\
            &=\frac{1}{8\pi G}\eps k_a \bar u_b K^{ab}\\
            &=\frac{1}{8\pi G}\eps\phi\bigl(
            c_{14} n^{a}_{\phantom{a}b}-c_{123}\delta^a_b -
            c_{13}h^a_b\big)
            \nabla_a u^b,
\end{split}
\eeq
where $\eps$ is the volume-element of $B$, $h_b^a =
g^{ac}h_{cb}$, and we have written, e.g.~$c_{14}$ for 
$c_1 + c_4$.
Next, we evaluate the $\delta u^a$-dependent portion of $i_{k}
\T_{\AE}$, pulled-back to $B$:
\beq
   \frac{1}{8\pi G}\, \eps n_{ab} k^b K^{a}_{\phantom{a}c}\delta
   u^c =\frac{1}{8\pi G}\, \eps
    \frac{\delta\phi}{\phi}k_a \bar u_b K^{ab} = \frac{\delta \phi}{\phi} \Q_{\AE}[k].
\eeq
The portion containing metric variations gives
\beq\label{ODD}
    \frac{1}{16\pi G}\,\eps  n_{ab}k^a u^b \delta h_{cd} K^{cd}
    =\frac{\phi}{16\pi G}\,\eps \frac{\delta A}{A}\big(
    c_{13}h^a_b - 2c_2\delta^a_b\big)\nabla_a u^b.
\eeq
We thus have
\begin{multline}\label{BADLAW}
    \int_{B} (\delta \Q_[\chi] - i_{\chi} \T)
        = \frac{\kappa}{8 \pi G}\bigg[\bigg(1
            +\phi\Big(c_{14}n^a_{\phantom{a}b} - c_{13}(\delta^a_b+\frac{3}{2}h^a_b)
            \Big)\nabla_a u^b\bigg)\delta A\\
        + \phi A
        \delta\Big((c_{14}n^a_{\phantom{a}b}-c_{123}\delta^a_b
                -c_{13}h^a_b\big)\nabla_a
                u^b\Big)\bigg],
\end{multline}
%
and we can then write the variational identity~\eqref{VARID} in
the form~\eqref{LAW}.

Although we have obtained this first-law \textit{form}, a
thermodynamic interpretation of it has not emerged.  In
particular, we do not have a definitive expression for the horizon
entropy.  For variations between members of a one-parameter family of
solutions, the horizon terms~\eqref{BADLAW} must be reducible to
$(\alpha\kappa/2\pi G) \delta A$ for some dimensionless constant 
$\alpha$.  
Even with the Noether charge approach, however, we still cannot  
discern the value of $\alpha$, nor do we know whether $\alpha A/G$ acts
as the entropy in the non-static case.  

It is possible that this confusion is related to an obscurity in
the notion of a black hole horizon in {\AE}-theory.  Linearized
perturbations about a `flat' background (flat space-time and
constant-{\ae}ther) were investigated in~\cite{Jacobson:2004ts}.
It was found that there exist five independent wave-modes that
travel at three different $c_i$-dependent speeds. These speeds
generally differ from the ``speed of light" defined by the flat
metric, and exceed it for certain $c_i$ values. The behavior of
perturbations about a curved background is not known, but we can
conjecture that a similar result holds.  If that is so, then a
Killing horizon is not generally a causal horizon.  On the other hand,
the
perturbations about the flat background do all propagate on the
light-cones of the flat metric~\cite{Jacobson:2004ts} in the
special case $c_{13} = c_{4} = 0$, $c_2 = c_1/(1-2c_1)$, but the
expression~\eqref{BADLAW} does not drastically simplify
in this case.  

Given the wide applicability of the principles of black hole
thermodynamics in generally covariant theories of gravity, it
would be surprising if they did not apply to {\AE}-theory. It seems
likely that the problem of formulating the laws of {\AE}-theory
black hole mechanics has a simple solution, which would
become apparent if we knew more about the explicit form of the solutions 
in the
neighborhood of the horizon.

%
%
\begin{acknowledgments}
I thank Robert Wald and Ted Jacobson for discussions, and the
Institut d'Astrophysique de Paris for hospitality while this work
was begun. This research was supported in part by the NSF under
grant PHY-0300710 at the University of Maryland.
\end{acknowledgments}
%
%
%

\end{document}